\documentclass[10pt, a4paper]{article}
\usepackage{lrec}
\usepackage{graphicx}
\usepackage{amsmath}
\usepackage{multirow}
\usepackage{url}
\usepackage{booktabs}
\usepackage[normalem,normalem,normalbf]{ulem}
\usepackage{tabularx}
\usepackage{soul}
\usepackage{tikz}
\usepackage[latin1]{inputenc}
\usepackage{xstring}
\usepackage{tikz}

\definecolor{FdWgreen}{rgb}{0.0, 0.42, 0.24}

\title{\textbf{Semi-supervised acoustic and language model training for English-isiZulu code-switched speech recognition}}

\name{A. Biswas, F. de Wet, E. van der Westhuizen, T.R. Niesler}

\address{Department of Electrical and Electronic Engineering, Stellenbosch University, South Africa \\
         \{abiswas, fdw, ewaldvdw, trn\}@sun.ac.za}
         
\abstract{
We present an analysis of semi-supervised acoustic and language model training for English-isiZulu code-switched ASR using soap opera speech.
Approximately 11 hours of untranscribed multilingual speech was transcribed automatically using four bilingual code-switching transcription systems operating in English-isiZulu, English-isiXhosa, English-Setswana and English-Sesotho.
These transcriptions were incorporated into the acoustic and language model training sets.
Results showed that the TDNN-F acoustic models benefit from the additional semi-supervised data and that even better performance could be achieved by including additional CNN layers.
Using these CNN-TDNN-F acoustic models, a first iteration of semi-supervised training achieved an absolute mixed-language WER reduction of 3.4\%, and a further 2.2\% after a second iteration.
Although the languages in the untranscribed data were unknown, the best results were obtained when all automatically transcribed data was used for training and not just the utterances classified as English-isiZulu.
Despite reducing perplexity, the semi-supervised language model was not able to improve the ASR performance.
\newline 
\Keywords{Code-switched speech, under-resourced languages, semi-supervised training, TDNN, CNN}
}

\begin{document}

\maketitleabstract
 
\section{Introduction}
\label{SEC:intro}

South Africa is a multilingual country with 11 official languages, including highly-resourced English which usually serves as a lingua-franca.
The largely multilingual population commonly mix these geographically co-located languages in casual conversation.
An ASR system deployed in this environment should therefore be able to process speech that includes two or more languages in one utterance. 
  
The study and development of code-switching speech recognition systems has recently attracted increased research attention~\cite{li2013improved,yilmaz2018semi,adel2015syntactic,emond2018transliteration}. 
Language pairs that are of current research interest include English-Mandarin \cite{li2013improved,vu2012first,zeng2018end}, Frisian-Dutch \cite{yilmaz2018semi,yilmaz2018building} and Hindi-English \cite{pandey2018phonetically}. 
In South Africa, code-switching most often occurs between highly resourced English and one of the nine under-resourced, officially-recognised African languages. 

In previous work, we showed that multilingual acoustic model training is effective for English-isiZulu code-switched ASR if additional training data from closely related languages is used~\cite{biswas2018IS}. 
However, the 12.2 hours of training data provided by combining all our code-switching data is still too little to  develop robust ASR systems. 

A related study indicated that increasing the pool of in-domain training data using semi-supervised training achieved a significant improvement over the baseline acoustic model~\cite{biswas2019IS2}.
These findings motivated us to further optimise  semi-supervised acoustic and language modelling training. Specifically, the effect of multiple iterations of semi-supervised training along with the application of a confidence threshold to filter the semi-supervised data was considered.
We focus our investigation on one language pair, English-isiZulu, to allow for a detailed analysis of various aspects of the semi-supervised training despite the limited computational resources at our disposal.
 

\section{Multilingual soap opera corpus} 
\label{SEC:corpus}

The multilingual speech corpus was compiled from 626 South African soap opera episodes.
Speech from these soap operas is typically spontaneous and   fast, rich in code-switching and often expresses emotion, making it a challenging corpus for ASR development. 
The data contains examples of code-switching between South African English and four Bantu languages: isiZulu, isiXhosa, Setswana and Sesotho.

\subsection{Manually transcribed data}
\label{SEC:corpus:transcribed}

Four language-balanced sets, transcribed by mother tongue speakers, were derived from the soap opera speech~\cite{van2018city}.
In addition, a large but language-unbalanced (English dominated) dataset containing 21.1 hours of code-switched speech data was created~\cite{biswas2019IS2}. 
The composition of this larger but unbalanced corpus is summarised in Table~\ref{tab:duration_unbalanaced_corpora}. 
Note that all utterances in the development and test sets contain code-switching and that the balanced data is a subset of the unbalanced data.

\begin{table}[h]
\footnotesize

\label{tab:duration_unbalanaced_corpora}
\centering 
\renewcommand*{\arraystretch}{1.0}
\resizebox{\columnwidth}{!}{
\begin{tabular}{@{}llrrrrr@{}}
\toprule
\multicolumn{1}{c}{}            & \bf{Language} & \bf{Mono} (m) & \bf{CS} (m) & \bf{Subtotal} (m) & \bf{Word tokens}             & \bf{Word types}             \\ \midrule
\multirow{5}{*}{\textit{Train}} & English  & 755.0   & 121.8 & 876.6     & 194~426 & 7~908  \\
                                & isiZulu  & 92.8    & 57.4  & 150.0     & 24~412  & 6~789  \\
                                & isiXhosa & 65.1    & 23.8  & 88.8      & 13~825  & 5~630  \\
                                & Sesotho  & 44.7    & 34.0  & 78.6      & 22~226  & 2~321  \\
                                & Setswana & 36.9    & 34.5  & 71.4      & 21~409  & 1~525  \\ \cmidrule(l){2-7} 
\textit{Dev}                    & EZ       & --       & 8.0    & 8.0       & 1~572   & 858                    \\ \cmidrule(l){2-7} 
\textit{Test}                   & EZ       & --       & 30.4   & 30.4      & 5~658   & 3711                   \\ \midrule
\bf{Total}                      &          & 994.4   & 271.5 & 1304.4    & 283~520 & 24~933 \\ \bottomrule
\end{tabular}%
}
\caption{Duration, in minutes (m), word type and word token counts for the unbalanced soap opera corpus. Both monolingual and code-switched (CS) durations are given.}
\end{table}

\vspace{-8pt}
\subsection{Manually segmented untranscribed data}
\label{SEC:corpus:untranscribed}

In addition to the transcribed data introduced in the previous section, 23\,290 segmented but untranscribed soap opera utterances were generated during the creation of the multilingual corpus. 
These utterances correspond to 11.1 hours of speech from 127 speakers (69 male; 57 female).
The languages in the untranscribed utterances are not labelled. Several South African languages not among the five present in the transcribed data are known to occur in these segments.


\section{Semi-supervised training}
\label{sec:sst}

\begin{figure*} [t]

	\centering
	\includegraphics[scale=0.05]{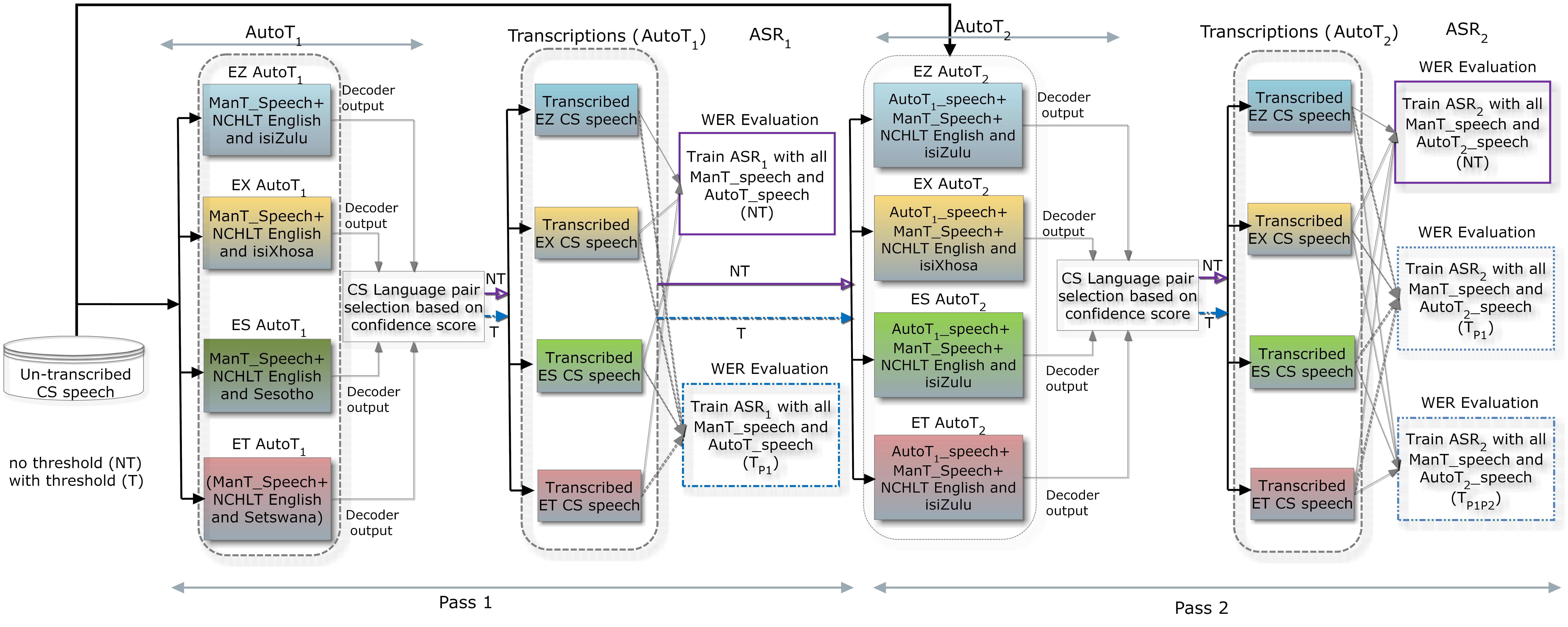}

 	\caption{Semi-supervised training framework for English-isiZulu code-switched (CS) ASR.}
		\label{fig_sst_scheme}

\end{figure*}

Semi-supervised techniques were used to transcribe the data introduced in Section~\ref{SEC:corpus:untranscribed} \cite{yilmaz2018semi,nallasamy2012semi,thomas2013deep}, starting with our best existing code-switching speech recognition system.
In this study the manually-segmented data was transcribed twice, as illustrated in Figure~\ref{fig_sst_scheme}. 
After each transcription pass, the acoustic models were retrained and recognition performance was evaluated in terms of WER. 

We distinguish between the acoustic models used to transcribe data ({\texttt{AutoT}}) and those that were used to evaluate WER (\texttt{ASR}) on the test set introduced in Table~\ref{tab:duration_unbalanaced_corpora}.
These two models differ in the composition of their training sets.

The acoustic models indicated by {\texttt{AutoT$_1$}} in Figure~\ref{fig_sst_scheme} were trained on all the manually transcribed (ManT) data described in Section~\ref{SEC:corpus:transcribed} as well as monolingual data from the NCHLT Speech Corpus~\cite{barnard2014nchlt}.
These were the best available models to start semi-supervised training.
The ManT and NCHLT data were subsequently pooled with the transcriptions produced by the \texttt{AutoT$_1$} models to train an updated set of acoustic models ({\texttt{AutoT$_2$}} in Figure~\ref{fig_sst_scheme}) which were used to obtain a new set of transcriptions of the untranscribed data for semi-supervised training.
In contrast, the acoustic models {\texttt{ASR$_1$}} and {\texttt{ASR$_2$}} were trained by pooling only the ManT and AutoT soap opera data; no out-of-domain NCHLT data was used.

Separate \texttt{AutoT} and \texttt{ASR} acoustic models are maintained because we use only in-domain data for semi-supervised training. 
This is computationally much easier, since the out-of-domain NCHLT datasets are approximately five times larger than the in-domain sets. 
However, it was found that better performance can be achieved in the second pass of semi-supervised training if the acoustic models maintain a similar training set composition to  that used in the first pass. 
Hence, \texttt{AutoT$_1$} and \texttt{AutoT$_2$} were purpose-built, intermediate systems used solely to generate semi-supervised data.

Figure~\ref{fig_sst_scheme} also shows that each untranscribed utterance was decoded by four bilingual ASR systems.
The highest confidence score was used to assign a language pair label to an utterance.
In initial experiments, we added only EZ data identified in this way to the pool of multilingual training data.
However, it was found that better performance could be achieved when all the AutoT data was added, and this was therefore done in the experiments reported here.
 
Two ways of augmenting the acoustic model training set with automatically-transcribed data were considered. 
First, all automatic transcriptions were pooled with the manually-labelled data.
Second, utterances with a recognition confidence score below a threshold were excluded.  
The average confidence score across each language pair was used as a threshold.
A larger variety of thresholds was not considered for computational reasons, but this remains part of ongoing work.
Confidence thresholds were applied in three ways.

\begin{enumerate}
    \item No threshold applied in either iteration~1 or~2 of semi-supervised training. The ManT data (21.1 h) was pooled with the AutoT$_1$ data to train {\texttt{ASR$_1$}} and with the AutoT$_2$ data to train {\texttt{ASR$_2$}}. The duration of both AutoT$_1$ and AutoT$_2$ was 11.1~h. 

    \item Threshold applied only in iteration~1. In this case only a subset of the AutoT$_1$ data (4.2 h) was pooled with the ManT data to train {\texttt{ASR$_1$}}. All 11.1~h of AutoT$_2$ data was used to train {\texttt{ASR$_2$}}.
    
    \item Threshold applied in both iteration~1 and iteration~2. This resulted in a 4.2 h subset of AutoT$_1$ used to train {\texttt{ASR$_1$}} and a 4.3 h subset of AutoT$_2$ used to train {\texttt{ASR$_2$}}.

\end{enumerate}

These three scenarios are indicated by ${NT}$, ${T_{P1}}$ and ${T_{P1P2}}$ respectively in Table~\ref{tab:AutoT}, which shows the number of utterances assigned to each language pair.
The total number of utterances and corresponding duration of the data included in the training set is shown in the last column.

\begin{table}[th]

 \footnotesize
  
  \label{tab:AutoT}
  \centering
  \renewcommand*{\arraystretch}{1.0}
  \resizebox{\columnwidth}{!}{%
  \begin{tabular}{ c c r r r r r}
    \toprule
    {\textbf{Pass}} & & {\textbf{EZ}} & {\textbf{EX}} & {\textbf{ES}} & {\textbf{ET}} & {\textbf{TOTAL}} \\
    \midrule
    1 & \multirow{2}{*}{{\textbf{${NT}$}}}   & 7\,951   & 3\,796 & 11\,415    & 128 & 23\,290 (11.1 h)\\
    2 &  & 9\,347   & 2\,145 & 5\,415     & 6\,381 & 23\,290 (11.1 h) \\
    \midrule
    1 & \multirow{2}{*}{{\bf{${T_{P1}}$}}}   & 3\,704   & 1\,731 & 5\,338   & 58 & 10\,831 (4.2 h)\\
    2 &  & 7\,888   & 1\,756 & 8\,798  & 4\,869 & 23\,290 (11.1 h)\\
    \midrule
    1 & \multirow{2}{*}{{\bf{${T_{P1P2}}$}}}   & 3\,704   & 1\,731 & 5\,338   & 58 & 10\,831 (4.2 h)\\
    2 &  & 3\,686 & 834  & 4\,115  & 2\,320 & 10\,955 (4.3 h)\\
    \bottomrule
  \end{tabular}
  }
  \caption{Number of utterances assigned to each language pair for automatically transcribed (AutoT) data.}
\end{table}


\section{Experiments}

\subsection{Language modelling}
\label{SEC:lm}

The English-isiZulu vocabulary consisted of 11\,292 unique word types and was closed with respect to the training, development and test sets.
The SRILM toolkit \cite{stolcke2002srilm} was used to train a bilingual trigram language model (LM) using the transcriptions described in Section~\ref{SEC:corpus:transcribed}
This LM was interpolated with two monolingual trigrams trained on 471 million English and 3.2 million isiZulu words of newspaper text, respectively.
The interpolation weights were chosen to minimise the development set perplexity.
The resulting language model was further interpolated with LMs derived from the transcriptions produced by the process illustrated in Figure~\ref{fig_sst_scheme} to obtain a semi-supervised LM.
 
\subsection{Acoustic modelling}
\label{SEC:am}

All ASR experiments were performed using the Kaldi toolkit \cite{povey2011kaldi} and the data described in Section~\ref{SEC:corpus} 
The automatic transcription systems were implemented using factorized time-delay neural networks (TDNN-F)~\cite{povey2018}.
For multilingual training, the training sets of all four language pairs were combined.
However, the acoustic models were language dependent and no phone merging across languages took place.
 
A context-dependent GMM-HMM was trained to provide the alignments for neural network training.
Three-fold data augmentation was applied prior to feature extraction~\cite{ko2015audio} and the acoustic features comprised 40-dimensional MFCCs (without derivatives), 3-dimensional pitch features and 100-dimensional i-vectors for speaker adaptation.

We used two types of neural network-based acoustic model architectures: (1) TDNN-F with 10 time-delay layers followed by a rank reduction layer  trained using the Kaldi Librispeech  recipe (version 5.2.164) and 
(2) CNN-TDNN-F consisting of two CNN layers followed by the TDNN-F architecture. 
TDNN-F models have been shown to be effective in under-resourced scenarios~\cite{povey2018}. The locality, weight sharing and pooling properties of the CNNs have been shown to benefit ASR~\cite{abdel2014convolutional}.
The default recipe parameters were used during neural network training.
In a final training step the multilingual acoustic models were adapted with English-isiZulu code-switched speech.


\section{Results and Discussion}

\subsection{Language modelling}

Table~\ref{TAB:perplexity} shows the test set perplexities (PP)  for the LM configurations described in Section~\ref{SEC:lm}
The baseline language model, LM$_0$, was trained on the English-isiZulu acoustic training data transcriptions as well as monolingual English and isiZulu text~\cite{biswas2018improving}.
LM$_0$ was also interpolated with trigram LMs trained on the 1-best and 10-best outputs of AutoT$_2$ respectively.
MPP indicates monolingual perplexity and is calculated over monolingual stretches of text only, omitting points at which the language alternates.
CPP indicates code-switch perplexity and is calculated only over language switch points.
Therefore CPP indicates the uncertainty of the first word following a language switch.

\begin{table}[h]
\footnotesize

\renewcommand*{\arraystretch}{1.0}
\label{TAB:perplexity}
\centering
\resizebox{\columnwidth}{!}{%
\begin{tabular}{l c c c c c c}
\toprule
LM & PP (dev) & PP  &  MPP$_E$ &  MPP$_Z$ &  MPP &  CPP \\ 
\midrule
LM$_{0}$ (baseline)           & 425.8 & 601.7  & 121.2 & 777.8 & 358.1  & 3\,292.0 \\
LM$_{0}$ + 1-best  & 416.1 & 587.4  & 123.1 & 743.6 & 351.1  & 3\,160.3 \\ 
LM$_{0}$ + 10-best & 408.2 & 583.6  & 124.4 & 722.8 & 346.9  & 3\,205.2 \\
\bottomrule
\end{tabular}%
}
\caption{Perplexity of bilingual English-isiZulu trigram LMs.}
\end{table}

Table \ref{TAB:perplexity} shows that, relative to the baseline, adding automatically generated English-isiZulu transcriptions to the language model training data improves the overall perplexity for both the development and test sets. 
The per-language results show that this improvement is due to a lower isiZulu perplexity, while English suffers a small deterioration.
CPP is reduced when incorporating the 1-best automatic transcriptions but less so when incorporating the 10-best.
This indicates that the code-switches present in the 1-best outputs are more representative of the unseen test set switches than those present in the 10-best output.

\subsection{Acoustic modelling}

ASR performance was evaluated  on the English-isiZulu test set for various configurations of the {\texttt{ASR$_1$}} and {\texttt{ASR$_2$}} systems.

\subsubsection{{\texttt{ASR$_1$}}}

Table~\ref {TAB:wer:pass1} reports WER results for different configurations of {\texttt{ASR$_1$}}.
Previously-reported results using a balanced subset of the corpus described in Section~\ref{SEC:corpus:transcribed} are reproduced in rows 1 and 2.
Language specific WERs are provided for the test set but not the development set.

The results in row 4 of the table show that, when the TDNN-F network is preceded by two CNN layers, test set recognition performance improves by 1.9\% absolute.
Row 5, on the other hand, shows that the inclusion of the automatically-transcribed English-isiZulu utterances reduces the test set WER of the TDNN-F models by 1.8\% absolute.
This improvement increases by an additional 0.8\% absolute when including all the automatically transcribed data and not just the English-isiZulu utterances, as shown in row 6.
Row 7 shows that the performance of the CNN-TDNN-F system is also enhanced by including the automatically transcribed data. In all the above cases, the WER improvements are seen not only overall but also in the English and isiZulu language-specific error rates.

Finally, the results in row 8 illustrate the impact of applying a confidence threshold to decide which automatically-transcribed utterances to include in the training set.
The values in the table indicate that the mixed WER deteriorates marginally and that the English WER improves at the cost of a higher isiZulu WER.

\begin{table}[h]
\footnotesize

\label{TAB:wer:pass1}
\centering
\renewcommand*{\arraystretch}{1.0}
\resizebox{\columnwidth}{!}{%
\begin{tabular}{c l r r r r }
\toprule
& \multicolumn{1}{l}{System configuration} & Dev & Test & \begin{tabular}[c]{@{}l@{}}WER$_E$\end{tabular} & \begin{tabular}[c]{@{}l@{}}WER$_Z$\end{tabular} \\ \midrule
1 & \begin{tabular}[c]{@{}l@{}}ManT (balanced)\\ [1mm]  TDNN-LSTM~\cite{biswas2018IS}\end{tabular} & 47.4 & 55.8 & 50.0 & 60.1 \\[1mm] 
2 & \begin{tabular}[c]{@{}l@{}}ManT (balanced) \\  TDNN-BLSTM~\cite{biswas2018improving}\end{tabular} & 47.1 & 53.1 & 47.6 & 57.2 \\ \midrule
3 & \begin{tabular}[c]{@{}l@{}}ManT (baseline) \\ TDNN-F\end{tabular} & 41.3 & 47.4 & 41.8 & 51.8 \\[1mm] 
4 & \begin{tabular}[c]{@{}l@{}}ManT \\ CNN-TDNN-F\end{tabular} & 40.8 & 45.6  & 40.0 & 49.9   \\  \midrule
5 & \begin{tabular}[c]{@{}l@{}}ManT + AutoT$_1$ (EZ,$NT$) \\ TDNN-F\end{tabular} & 41.2 & 45.7 & 39.6 & 50.3 \\[1mm]  
6 & \begin{tabular}[c]{@{}l@{}}ManT + AutoT$_1$ (All,$NT$) \\ TDNN-F \end{tabular} & 39.5 & 44.9 & 38.9 & 49.6 \\  \midrule
7 & \begin{tabular}[c]{@{}l@{}}ManT + AutoT$_1$ (All,$NT$) \\ CNN-TDNN-F \cite{biswas2019IS2} \end{tabular} & {\textbf{38.2}} & {\textbf{44.0}} & 37.9 & 48.7 \\[1mm]  
8 & \begin{tabular}[c]{@{}l@{}}ManT + AutoT$_1$ (All,$T_{P1}$) \\ CNN-TDNN-F \end{tabular} & 38.8 & 44.2 & 36.6 & 50.1 \\[1mm]  
\bottomrule
\end{tabular}%
}
\caption{WER (\%) on the English-isiZulu development (dev) and test sets for different configurations of {\texttt{ASR$_1$}}.}
\end{table}

\subsubsection{{\texttt{ASR$_2$}}}

The results for the second iteration of semi-supervised training are reported in Table \ref{TAB:wer:pass2}.
In all cases the ManT data was pooled with all the AutoT data and not just the EZ sub-set as was done in row 5 of Table~\ref {TAB:wer:pass1}.
Only the results using the CNN-TDNN-F acoustic models are shown, since this gave consistently superior performance in Table~\ref {TAB:wer:pass1}.

\begin{table}[h]
\footnotesize

\label{TAB:wer:pass2}
\resizebox{\columnwidth}{!}{%
\begin{tabular}{@{}lllllll@{}}
\toprule
 & Training data & LM & Dev & Test & WER$_{E}$ & WER$_{Z}$ \\ \midrule
1 & \begin{tabular}[c]{@{}l@{}}ManT + AutoT$_2$\\ (NT)\end{tabular} & LM$_{0}$ & 38.6 & 42.5 & 36.2 & 47.6 \\
2 & \begin{tabular}[c]{@{}l@{}}ManT + AutoT$_2$\\ ($T_{P1}$)\end{tabular} & LM$_{0}$ & 38.0 & 43.1 & 37.5 & 47.4 \\
3 & \begin{tabular}[c]{@{}l@{}}ManT + AutoT$_2$\\ ($T_{P1P2}$)\end{tabular} & LM$_{0}$ & 40.1 & 43.9 & 34.2 & 51.3 \\ \midrule
4 & \multirow{3}{*}{\begin{tabular}[c]{@{}l@{}}ManT + AutoT$_2$\\ (NT, tuned)\end{tabular}} & LM$_{0}$ & 36.5 & 41.9 & 33.0 & 48.8 \\
5 &  & LM$_{0}$ + 1-best & \bf{36.5} & \bf{41.8} & 33.9 & 47.9 \\
6 &  & LM$_{0}$ + 10-best & 36.7 & 42.0 & 34.0 & 48.1 \\ \bottomrule
\end{tabular}%
}
\caption{WER (\%) on the English-isiZulu development (dev) and test sets for different configurations of {\texttt{ASR$_2$}}.}
\end{table}

A comparison between row 1 in Table~\ref{TAB:wer:pass2} and row 7 in Table~\ref {TAB:wer:pass1} reveals that a second pass of retraining affords a further 1.5\% absolute reduction in test set WER.
This was found to be statistically significant at more than 95\% confidence level using bootstrap interval estimation \cite{bisani2004bootstrap}.
Retraining {\texttt{ASR$_2$}} with a threshold applied only to the output of AutoT$_1$ results in a slightly higher WER on the test set (row 2). 
Applying thresholds in both passes (row 3) improved the English WER but resulted in a substantial deterioration in isiZulu WER. 
This result suggests that, for the threshold value used here, English benefits from the exclusion of low-confidence automatically transcribed data while isiZulu does not. 
Thus, further study on the optimum threshold configuration is required.

The results in row 4 of Table~\ref{TAB:wer:pass2} show that a further 0.6\% absolute WER reduction can be achieved for the test set by tuning the learning rate during adaptation.
Rows 5 and 6 show that retraining the LM on text that includes automatic transcriptions hardly influences recognition performance. 
Thus, although semi-supervised  training led to appreciable improvements in the acoustic models, the  corresponding positive effects on the language model were marginal.

A detailed analysis of different ASR outputs is shown in Table \ref{detailed_output}. The analysis confirms that semi-supervised training resulted in substantial improvements in the English and isiZulu word correct accuracy. The results also reveal a substantial improvement in bigram correct accuracy at the 1\,464 code-switch points occurring in the test set, where bigram correct accuracy (\%) is defined as the percentage of words correctly recognised immediately after code-switch points.

\begin{table}[h]

\label{detailed_output}
\centering
\resizebox{\columnwidth}{!}{%
\begin{tabular}{ l l l l l l}
\toprule
  Accuracy (\%)  & 
 \begin{tabular}[c]{@{}l@{}}Table 4\\ (Row 3)\end{tabular} & \begin{tabular}[c]{@{}l@{}}Table 4\\ (Row 4)\end{tabular} & \begin{tabular}[c]{@{}l@{}}Table 4\\ (Row 7)\end{tabular} & \begin{tabular}[c]{@{}l@{}}Table 4\\ (Row 8)\end{tabular} & \begin{tabular}[c]{@{}l@{}}Table 5\\ (Row 4)\end{tabular} \\ \midrule
Eng token correct & 59.8 & 61.5 & 64.5 & 65.4 & 68.8 \\ 
Zul token correct & 50.1 & 51.4 & 53.2 & 51.6 & 53.5 \\ 
Word correct after switch & 53.4 & 55.6 & 58.3 & 57.6 & 60.9 \\ 
Zul word correct after switch & 49.7 & 51.4 & 53.6 & 51.6 & 54.4 \\ 
English word correct after switch & 56.7 & 59.3 & 62.5 & 62.9 & 66.7 \\ 
Language correct after switch & 76.8 & 76.9 & 79.1 & 79.0 & 81.6 \\ 
Code-switch bigram correct & 29.0 & 30.8 & 33.3 & 32.2 & 35.6 \\
\bottomrule
\end{tabular}%
}
\caption{Detailed analysis of ASR accuracy for different acoustic models.}
\end{table}

\section{Conclusion}

We have applied semi-supervised training to improve ASR for under-resourced code-switched English-isiZulu speech. 
Four different automatic transcription systems were used in two phases to decode 11 hours of multilingual, manually segmented but untranscribed soap opera speech. 
We found that by including CNN layers, CNN-TDNN-F acoustic models outperformed TDNN-F models on the code-switched speech.
Furthermore, semi-supervised training provided a further absolute reduction of 5.5\% in WER for the CNN-TDNN-F system. 
While the automatically transcribed English-isiZulu text data reduced language model perplexity, this improvement did not lead to a reduction in WER. 
By selective data inclusion using a confidence threshold, approximately 60\% of the automatically transcribed data could be discarded at minimal loss in recognition performance.
A more thorough investigation of this threshold remains part of ongoing work.
We also aim to further extend the pool of training data by incorporating speaker and language diarisation systems to allow automatic segmentation of new audio.

\section{Acknowledgements}
We would like to thank the Department of Arts \& Culture (DAC) of the South African government for funding this research. 
We are grateful to e.tv and Yula Quinn at Rhythm City, as well as the SABC and Human Stark at Generations: The Legacy, for assistance with data compilation. 
We also gratefully acknowledge the support of the NVIDIA corporation for the donation of GPU equipment.

\section{References}

\bibliographystyle{lrec}



\end{document}